# Homogeneity tests for Michaelis-Menten curves with application to fluorescence resonance energy transfer data

**Amparo Baíllo[1*], Laura Martínez-Muñoz[2], Mario Mellado[2]**

[1]Departamento de Matemáticas, Universidad Autónoma de Madrid, Ciudad Universitaria de Cantoblanco, 28049 Madrid, Spain

[2]Department of Immunology and Oncology, Centro Nacional de Biotecnología/CSIC, Ciudad Universitaria de Cantoblanco, 28049 Madrid, Spain

[*]Corresponding author

E-mail addresses:

    AB: amparo.baillo@uam.es

    LMM: lmmunoz@cnb.csic.es

    MM: mmellado@cnb.csic.es

# Abstract

**Background**

Resonance energy transfer (RET) methods are in wide use for evaluating protein-protein interactions and protein conformational changes in living cells. Sensitized emission fluorescence resonance energy transfer (FRET) measures energy transfer as a function of the acceptor-to-donor ratio, generating FRET saturation curves. Modeling the saturation curves by Michaelis-Menten kinetics allows characterization by two parameters, $FRET_{max}$ and $FRET_{50}$. These parameters allow evaluation of apparent affinity between two proteins and comparison of this affinity in different experimental conditions. To reduce the effect of sampling variability, several replications (statistical samples) of the saturation curve are generated in the same biological conditions. Here we study procedures to determine whether statistical samples in a collection are homogeneous, in the sense that they are extracted from the same underlying saturation curve (or regression model).

**Results**

We used three methods to determine which statistical samples in a group are homogeneous. From the hypothesis testing viewpoint, we considered two procedures: one based on bootstrap resampling and the other, a version of a classical $F$ test. The third method analyzed the problem from the model selection viewpoint, and used the Akaike information criterion (AIC). Although we only considered the Michaelis-Menten model, all three statistical procedures would also be applicable to any other nonlinear regression model. We compared the performance of the three homogeneity testing methods in a Monte Carlo study and through analysis in living cells of FRET saturation curves for dimeric complexes of CXCR4, a seven-transmembrane receptor of the G protein-coupled receptor family.

**Conclusions**

The simulation study and analysis of real FRET data showed that the *F* test, the bootstrap procedure and the model selection method lead in general to similar conclusions, although AIC gave the best results when sample sizes were small, whereas the *F* test and the bootstrap method were more appropriate for large samples. In practice, all three methods are easy to use simultaneously and show consistency, facilitating conclusions on sample homogeneity.

# Background

Oligomerization, the formation of a complex by two or more proteins, is a subset of protein-protein interactions that generates considerable functional diversity. It frequently operates in the transduction of signals that begin at the cell surface and continue to the nucleus, in pathways that participate in antigen receptor signaling, cytokine responses, regulation of gene transcription, and in so many other processes that it clearly constitutes a major mechanism in the regulation of cell responses.

The classical biochemical methods for monitoring these interactions include coimmunoprecipitation and western blot studies using untagged or tagged proteins, or crosslinking analysis, which uses solubilizing detergents and disrupted cells; both can introduce artifacts. Difficulty increases when we evaluate proteins with a complex structure at the cell membrane, such as the G protein-coupled receptors (GPCR). This protein family is characterized by seven $\alpha$-helical domains that span the cell membrane, and is one of the most abundant in nature. It includes receptors for hormones, neurotransmitters, chemokines and calcium ions, among others, and is thus a focal point of the pharmaceutical industry's effort to develop antagonists for therapeutic use in man.

New methods based on resonance energy transfer (RET) are becoming widespread for the evaluation of protein-protein interactions in living cells. These techniques can also be used to define protein rearrangement, conformation dynamics or the role of ligand and receptor levels, to screen for antagonists, and to study dimerization sites within the cell [1]. There are two main types of RET, fluorescence resonance energy transfer (FRET) and bioluminescence resonance energy transfer (BRET); in the former, the donor is a fluorescent protein that transfers energy to an acceptor fluorescent molecule and in the latter, the donor molecule is luminescent [2, 3].

FRET is based on non-radiative energy transmission from a donor fluorophore to a nearby acceptor, without photon emission. Energy transfer depends on the overlap between the donor emission spectrum and acceptor absorption, on the distance between donor and acceptor (which must be in the 2-10 nm range), and requires correct orientation between donor and acceptor fluorophores [4, 5]. FRET determinations use the protein of interest fused to distinct spectral variants of the green fluorescent protein (GFP); the most commonly used variants are cyan (CFP) as donor and yellow (YFP) as acceptor [6, 7].

Sensitized emission FRET allows measurement of energy transfer in reference to the acceptor-to-donor ratio to generate FRET saturation curves. These curves describe FRET efficiency as a function of the acceptor-to-donor ratio and are characterized by two important parameters. The first, $B_{max}$ (also frequently denoted $FRET_{max}$), is the (asymptotic) maximum of the curve. If energy transfer reaches saturation and the curve is hyperbolic, we define a second parameter, usually denoted $K_d$ or $FRET_{50}$, that corresponds to the acceptor-to-donor ratio that yields half $FRET_{max}$ efficiency. The parameter $K_d$ allows estimation of the apparent affinity between the partners involved [8, 9]. Both parameters depend on the distance between donor and acceptor, and on their orientation in the complex; $B_{max}$ and $K_d$ are thus directly related to the energy transferred, and therefore to the number of protein complexes formed and/or to changes in complex conformations. Although here we have only considered FRET curves, the same analysis and conclusions are applicable to BRET titration.

$B_{max}$ and $K_d$ allow evaluation of oligomerization of two proteins (dimerization) in different experimental conditions. For example, they can be used to assess how the presence of a third coexpressed protein affects complex conformation. In practice, $B_{max}$ and $K_d$ are estimated using a statistical sample of points $(x_i, y_i)$, $i = 1,\ldots,n$, from the saturation curve. To avoid misunderstanding, we note that when we refer to "sample" in the statistical sense, we refer to that which in biology is termed a "saturation curve". To reduce the effect of sampling variability on the comparison of two distinct experimental conditions, several replications (or statistical samples) of the saturation curve are usually generated in each condition.

For this study, we used seven statistical samples $(x_{ij}, y_{ij})$, $i = 1,\ldots,7$, $j = 1,\ldots,n_i$ of a FRET saturation curve for CXCR4 dimers in living cells (Figure 1; see Results and Discussion for more information on the data). CXCR4 is a chemokine receptor of the GPCR family, with key roles in homeostasis and pathology. Mice lacking CXCR4 die perinatally and have defects in vascular development, hematopoiesis and cardiogenesis [10]. CXCR4 is also implicated in cancer [11], rheumatoid arthritis [12] and pulmonary fibrosis [13]; finally, together with CCR5, CXCR4 is one the main coreceptors for HIV-1 infection [14].

Statistical analysis is necessary to determine which samples are homogeneous, in the sense that they are extracted from the same underlying saturation curve (regression model). The homogeneous samples from each saturation curve will be those finally considered for comparison of distinct experimental conditions, e.g., alone or in the presence of an additional protein.

Our goal was to define a method for reliable comparison of protein dimerization before and after a specific change in experimental conditions (statistically also termed "treatments"). If there is only one version or statistical sample of each distinct experiment, it is reasonable to use any of the procedures (a *t* test, for instance) described by Motulsky & Christopoulos [15]. In other cases, several statistical samples are obtained of the experiment before and after the change, which gives rise to several estimated FRET curves in the two experimental conditions.

Curves in these two groups are sometimes naturally paired, for example, when dimerization is evaluated in the same group of cells before and after the change in experimental conditions. We then have $I$ pairs of saturation curves, and a $t$ test can be used to compare the components of each pair (for example, see [15-17]). If the majority of the $I$ $p$-values from the $t$ tests is <0.05, the conclusion is that the effect of the treatment is statistically significant.

When the curves before and after the treatment are not paired, it is reasonable to focus first on each of the two samples of curves separately. A sample of $I$ saturation curves represents $I$ realizations in the same experimental conditions, which (intuitively) should correspond to observations of *the same* probability model. One possible procedure is to fit two random-effects models, one to the experiment before the treatment and another to that after the treatment [15]. The realizations of the random effects in a specific model would account for the differences between versions of the same experiment (see [18] for the random-effects version of the Michaelis-Menten model (1), used to describe saturation curves). Although this idea is appealing, the appropriateness of fitting a random-effects model when the number of samples $I$ is low (e.g., $I = 3$) is questionable.

Here we propose to verify, via a homogeneity test of hypotheses, which of the $I$ realizations of an experiment can be accepted to come from the same Michaelis-Menten model. Statistical samples corresponding to homogeneous outcomes can then be pooled into a unique sample from the common underlying model. Once we have determined a homogeneous sample for each experiment (before and after the change), we can apply a $t$ or an $F$ test to determine whether there are differences due to the change in experimental conditions. For the homogeneity test, we consider two different testing procedures (an $F$ test and a resampling-based scheme) and compare their behavior via a simulation study and the analysis of real FRET data; in all cases, we also computed the Akaike information criterion (AIC) for the models implied by the null and alternative hypotheses. Although the theories underlying the information-theoretic approach and null-hypothesis testing differ [19], the conclusions derived from all the approaches are in general the same.

## Methods

**Michaelis-Menten model**

The saturation curve, that is, FRET efficiency ($Y$) as a function of the acceptor-to-donor ratio ($X$), is usually described via the nonlinear regression model of Michaelis-Menten [5, 15, 20, 21]

$$Y = \frac{B_{max} X}{K_d + X} + \sigma \varepsilon, \qquad (1)$$

where $\varepsilon$ follows a standard normal distribution. Throughout this study, we estimate the unknown parameters $\theta = (B_{max}, K_d)$ and $\sigma$ by maximum likelihood. For a sample $(x_i, y_i)$, $i = 1,\ldots,n$, of independent observations of model (1), the maximum likelihood estimators (m.l.e.) are given by $\hat{\theta} = (\hat{B}_{max}, \hat{K}_d) = \arg\min_{\theta} S(\theta)$ and $\hat{\sigma}^2 = S(\theta)/n$, where $S(\theta) = \sum_{i=1}^{n}(y_i - B_{max} x_i /(K_d + x_i))^2$ is the residual sum of squares.

**The homogeneity test for Michaelis-Menten curves**

We have data $(x_{ij}, y_{ij})$, $i = 1,\ldots,I$, $j = 1,\ldots,n_i$, from $I$ realizations of an experiment. For each $i$, the sample $(x_{ij}, y_{ij})$, $j = 1,\ldots,n_i$, is assumed to be observed from the model

$$Y = \frac{B_{max;i} X}{K_{d;i} + X} + \sigma \varepsilon, \qquad (2)$$

where $\varepsilon$ follows a standard normal distribution.

We wish to test whether the data from the $I$ samples are observations from the same model of type (1), that is, we are interested in the hypothesis test

$$H_0: \quad \theta_1 = \theta_2 = \ldots = \theta_I \; (= \theta_0) \qquad (3)$$

$$H_1: \quad \theta_i \neq \theta_j \text{ for some } j \neq k$$

where $\theta_i = (B_{max;i}, K_{d;i})$. Accepting $H_0$ means that all the observations $(x_{ij}, y_{ij})$, $i = 1, \ldots, I$, $j = 1, \ldots, n_i$ are outcomes of the same experiment and can be pooled into a sample of size $n = n_1 + \ldots + n_I$ to give a single estimation of $\theta_0$. This would be the desirable conclusion when the $I$ samples are observed in the same experimental conditions (as in Figure 1). Since the data are observations of protein-protein interactions in *live* cells, however, it is frequent that, due to uncontrollable factors, at least one of the $I$ statistical samples appears to be different from the majority (see, for example, sample 7 in Figure 1). As the final aim is to compare results in distinct experimental conditions, it is important first to decide which estimated saturation curves are homogeneous in the same experimental conditions.

**Rejection regions**

Let us first fix the following notation. Under $H_1$ the m.l.e. of $\theta_i$ is given by $\hat{\theta}_i = (\hat{B}_{max;i}, \hat{K}_{d;i}) = \arg\min_\theta S_i(\theta)$, where

$$S_i(\theta) = \sum_{j=1}^{n_i} \left( y_{ij} - \frac{B_{max} x_{ij}}{K_d + x_{ij}} \right)^2$$

and the m.l.e. of $\theta_0$ under $H_0$ is $\hat{\theta}_0 = (\hat{B}_{max;0}, \hat{K}_{d;0}) = \arg\min_\theta \sum_{i=1}^{I} S_i(\theta)$.

We consider two possible ways of constructing a rejection region for the test (3); one is based on a bootstrap resampling scheme [22] and the other is derived from an $F$ test.

*Bootstrap rejection region*

Let us consider the test statistic

$$T = \log\left(\sum_{i=1}^{I} S_i(\hat{\theta}_0)\right) - \log\left(\sum_{i=1}^{I} S_i(\hat{\theta}_i)\right) \tag{4}$$

proportional to the log-likelihood ratio. We reject $H_0$ when $T > T_{0;\alpha}$, the $(1-\alpha)$ quantile of $T$ under $H_0$. The value of $T_{0;\alpha}$ has been approximated via bootstrap, with the following algorithm:

1. For the original sample $(x_{ij}, y_{ij})$, $i = 1,\ldots,I$, $j = 1,\ldots,n_i$, compute the m.l.e. $\hat{\theta}_0$ and $\hat{\sigma}^2$ under $H_0$.

2. Fix a (typically large) number $B$ of bootstrap samples. In this study, we chose $B = 1000$.

3. For every $b = 1,\ldots,B$ draw a sample $y_{ij}^{(b)} = \hat{B}_{max;0} x_{ij} / (\hat{K}_{d;0} + x_{ij}) + \hat{\sigma}\varepsilon_{ij}^{(b)}$, $i = 1,\ldots,I$, $j = 1,\ldots,n_i$, where $\varepsilon_{ij}^{(b)}$ follows a standard normal distribution.

4. Compute $T^{(b)}$, the value of the test statistic (4), for the $b$-th bootstrap sample $(x_{ij}, y_{ij}^{(b)})$, $i = 1,\ldots,I$, $j = 1,\ldots,n_i$.

5. As an approximation to $T_{0;\alpha}$, take $\hat{T}_{0;\alpha}$, the $[(1-\alpha)B]$-th order statistic of the sample $T^{(1)},\ldots,T^{(B)}$, where $[a]$ denotes the least integer greater than or equal to $a$.

*F test*

There are several proposals of $F$ test statistics in the nonlinear regression literature (see [21] for a review). To apply these ideas to our problem, we reparameterize the model given in (2) as follows. We consider the global vector of parameters $\gamma = (\gamma_1,\ldots,\gamma_{2(I+1)})'$, where $\gamma_1 = B_{max;0}$, $\gamma_2 = K_{d;0}$, $\gamma_{2i+1} = B_{max;i} - B_{max;0}$, $\gamma_{2i+2} = K_{d;i} - K_{d;0}$, for $i = 1,\ldots,I$ and $\sum_{i=1}^{I} \gamma_{2i+j} = 0$ for $j = 1, 2$. Then the test given in (3) is equivalent to

$$H_0: \quad \gamma_3 = \ldots = \gamma_{2(I+1)} = 0 \tag{5}$$

$$H_1: \quad \gamma_k \neq 0 \text{ for some } k \geq 3.$$

We consider the test statistic

$$F = \frac{\sum_{i=1}^{I}\left(S_i(\hat{\theta}_0) - S_i(\hat{\theta}_i)\right)}{\sum_{i=1}^{I} S_i(\hat{\theta}_i)} \frac{n - 2I}{2(I-1)}, \tag{6}$$

which, under $H_0$ in tests (3) or (5), follows approximately an $F_{2(I-1); n-2I}$ distribution [21]. Consequently, we reject the null hypothesis of homogeneity when $F > F_{2(I-1); n-2I; \alpha}$.

*Model selection*

Between models (1) and (2), deciding which is the most appropriate to fit and analyze the information contained in the sample can also be viewed as a model selection problem [23]. In this case, we can use AIC to select the model best approximating the data [24]. The information criterion corresponding to the general model (2) is given by $AIC = n \log\left(\sum_{i=1}^{I} S_i(\hat{\theta}_i)/n\right) + 2(1 + 2I)$. The more restrictive model (1) is nested in this class, satisfies $\theta_1 = \ldots = \theta_I = \theta_0$ and its information criterion is $AIC_0 = n \log\left(\sum_{i=1}^{I} S_i(\hat{\theta}_0)/n\right) + 6$. To select the best model, we compute the AIC difference $\Delta = AIC_0 - AIC$. A small value of $\Delta$ (say $0 \leq \Delta \leq 2$) indicates that model (1) is better than (2). If $\Delta$ is large, then model (2) is to be preferred. We refer the reader to [19, 23] for more details on information criteria-based decisions.

We are aware that this information-theoretic paradigm and hypothesis testing are very different approaches to the problem at hand, and should not be mixed. We have nonetheless found that the AIC can serve to corroborate the decision of accepting or rejecting the homogeneity of Michaelis-Menten curves, especially in the analysis of real data. We consider it interesting to focus on the problem from this point of view and to notice that, in this study, the information criterion and hypothesis testing lead to similar conclusions.

## Results and Discussion

We compared the performance of the homogeneity testing procedures described above via a Monte Carlo study and analysis of FRET data obtained in the laboratory.

**Simulations**

We consider the homogeneity test (3). In this subsection, we describe and interpret the results of a simulation study carried out to compare the power of the testing procedures introduced above. In all cases, the significance level is $\alpha = 0.05$ and the number of Monte Carlo runs is 1000.

Let us first describe how the observations were generated. We fix the number $I$ of samples whose homogeneity we want to test, the values of $n_i$, $B_{max;i}$ and $K_{d;i}$, for $i = 1,\ldots,I$, and $\sigma$. Then, for each $i = 1,\ldots,I$, we generate $n_i$ independent observations $X$ from a uniform distribution on the interval $[0, 4K_{d;i}]$ and $n_i$ independent $\varepsilon$ from a N(0,1) distribution. The random variables $X$ and $\varepsilon$ are independent. The corresponding values of the response $Y$ are computed according to model (2).

For simplicity, in each simulation we fix equal sample sizes $n_1 = n_2 = \ldots = n_I$. We consider three values for $n_i$: 20, 50 and 100. To illustrate the case in which $H_0$ is true, we chose $I = 2$ and $I = 5$, $\theta = (B_{max}, K_d) = (0.75, 0.5)$, $\theta = (1, 1.5)$ and $\sigma = 0.01$ or 0.001. The results of the corresponding simulations are shown in Table 1. In the second and third columns from the right, we record the proportion of times that $H_0$ is rejected using the bootstrap rejection region and the $F$ test, respectively. The last column shows the proportion of times that the AIC difference, $\Delta$, is greater than 2. In the table, the proportion of $H_0$ rejections is always near the nominal value of $\alpha = 0.05$ or well below it. For $I > 2$, this proportion is much lower for the AIC-based method than for the bootstrap procedure or the $F$ test. When $B_{max} = 1$, the number of times that $H_0$ is rejected is lower for the $F$ test than for the bootstrap method. This agrees with the results for $B_{max;i} \approx 1$ shown in Table 2.

Simulation results on the proportion of $H_0$ rejections when the null hypothesis is false appear in Tables 2 and 3. We compared $I = 2$ and $I = 3$ curves (see Tables 2 and 3, respectively), since the case $I \geq 4$ involves choosing many different parameters, but does not provide any more useful information than $I = 3$. The sample sizes $n_i$ coincide with those of Table 1. Here we only used $\sigma = 0.01$, as $\sigma = 0.001$ yielded a considerable number of (less interesting) cases in which the power was nearly 1. The value of the parameters $\theta_i$ chosen for these simulations is in the neighborhood of those in Table 1.

The simulation results in Tables 2 and 3 show that, for $B_{max;i}$ near 0.75, the power of the AIC-based procedure is greater than that of the bootstrap method and the $F$ test when $n_i = 20$, for $i = 1, \ldots, I$; that is, for a small sample size. This is due to the facts that bootstrap is a resampling-based technique [22] and the $F$ test statistic (6) only follows Fisher's distribution asymptotically (for large sample sizes $n_i$) under the null hypothesis [21]. For larger sample sizes ($n_i = 50$ or 100) and $I = 3$, the bootstrap procedure and the $F$ test yield very similar powers, in general slightly better than those attained by AIC. When $B_{max;i}$ is near 1, the boundary of the parameter space for $B_{max}$, then the bootstrap procedure is superior to both other methods in all cases, as also happened when $H_0$ was true. In the case of the $F$ test, this is because the asymptotic distribution of the statistics involved (such as the m.l.e. of the parameters) is obtained under the restriction that the unknown parameters lie in the interior of the parameter space [21]. In FRET data analysis, this restriction is not important from a practical point of view, since values of $B_{\max} \geq 1$ do not occur. Values of $B_{\max} = 1$ would indicate that all energy emitted by the donor protein is absorbed by the acceptor protein. This is only the case when CFP-YFP are fused in tandem, as for a positive control. When CFP and YFP are fused to the C-terminal end of the GPCR, the theoretical distance between CFP-YFP (5.28 nm) and the orientation allowing 100% energy transfer between donor and acceptor are modified; real $B_{\max}$ is therefore always < 1 [25]. As the case $B_{\max;i} \approx 1$ is thus only interesting from a statistical point of view, Table 3 shows fewer simulations with $B_{\max;i} \approx 1$ than with $B_{\max;i} \approx 0.75$.

**FRET data**

In this subsection, we study the homogeneity of the *I* = 7 FRET statistical samples in Figure 1.

*Laboratory material and methodology*

HEK293T (human embryonic kidney) cells from the American Type Culture Collection (CRL-11268) were plated in 6-well plates (Nunc) 24 h before transfection (5 x $10^5$ cells/well). Cells were transiently transfected with cDNA encoding the fusion proteins (CXCR4-CFP or CXCR4-YFP) by the polyethylenimine method (PEI; Sigma-Aldrich). Cells were incubated with DNA and PEI (5.47 mM in nitrogen residues) and 150 mM NaCl in serum-free medium, which was replaced after 4 h with complete medium. At 48 h post-transfection, cells were washed twice in Hank's balance salt solution supplemented with 0.1% glucose, and resuspended in the same solution. Total protein concentration was determined for whole cells using a Bradford assay kit (BioRad). Cell suspensions (20 μg protein in 100 μl) were pipetted into black 96-well microplates and read in a Wallac Envision 2104 Multilabel reader (Perkin Elmer) equipped with a high-energy xenon flash lamp, using an 8 nm bandwidth excitation filter at 405 nm (393-403 nm) and 10 nm bandwidth emission filters at 486 nm (CFP channel) and 530 nm (YFP channel). Gain settings were optimized for each experiment to maintain a constant relative contribution of fluorophores to the detection channels for spectral imaging and linear unmixing. To determine the spectral signature, HEK293T cells were transiently transfected with CXCR4-CFP or CXCR4-YFP. The contributions of CFP and YFP alone were measured in each detection channel (spectral signature), and normalized to the sum of the signal obtained for both channels [17, 26, 27]. The spectral signatures of CXCR4-CFP and CXCR4-YFP did not vary significantly ($p > 0.05$) from the signatures determined for each fluorescent protein alone. For FRET quantitation, the spectral signature was taken into consideration for linear unmixing to separate the two emission spectra.

In each experimental condition, we thus measured FRET efficiency for multiple acceptor-to-donor ratios that are obtained by maintaining a constant donor amount and increasing amounts of acceptor. To quantify acceptor-to-donor ratio, we first determined the total amount of donor protein by excitation with its specific wavelength (405 nm) and measurement at 486 nm, as well as the total amount of acceptor protein by excitation at 515 nm and measurement at 530 nm. FRET efficiency was then calculated for each acceptor-to-donor ratio using the formulas CFP = S/(1+1/R) and YFP = S/1+R, where S = ChCFP + ChYFP, R = (YFP$_{530}$Q - YFP$_{486}$)/(CFP$_{486}$ - CFP$_{530}$Q) and Q = ChCFP/ChYFP. ChCFP and ChYFP represent the signal in the 486 nm and 530 nm detection channels (Ch). CFP$_{486}$, CFP$_{530}$, YFP$_{530}$, and YFP$_{486}$ represent the normalized contributions of CFP and YFP to channels 486 and 530, as determined from spectral signatures of the fluorescent proteins.

**Statistical analysis**

In each realization $i$ in the same experimental conditions, we obtain a statistical sample $(x_{ij}, y_{ij})$, $j = 1, \ldots, n_i$, where the response variable $Y$ represents FRET efficiency and the explanatory variable $x$ is the acceptor-to-donor ratio. Figure 1 shows $I = 7$ samples of the saturation curve corresponding to the experiment described above. In Table 4, we summarize statistical information on the data.

We study the homogeneity of these samples via the three test procedures described (see Methods). Since estimation of $\theta = (B_{max}, K_d)$ is highly sensitive to the presence of outliers, we used the ROUT outlier detection procedure [28] to identify and remove this type of data before further analysis. In Figure 2, we show the estimated Michaelis-Menten curves for the samples in Figure 1 after removal of the outlying observations. We marked the values of $\hat{B}_{max;i}$ and $\hat{K}_{d;i}$, for $i = 1,\ldots,7$, on the vertical and horizontal axes of Figure 2, respectively. The seventh curve clearly differs from the first six, which are almost identical to one another. In effect, the bootstrap and the $F$ test procedures reject the null hypothesis (3) of homogeneity for significance level $\alpha = 0.01$. In the bootstrap procedure, the test statistic (4) takes the value $T = 1.08$ and the critical value is $\hat{T}_{0;0.01} = 0.23$. In the $F$ test, the statistic (6) is $F = 16.5$ and the critical value is $F_{12;102;0.01} = 2.4$. The AIC difference is $\Delta = 100.94$, so according to AIC, we would choose the general model (2). The three procedures thus lead to the same conclusion: the seven curves are not homogeneous. Let us now remove the seventh curve and carry out the homogeneity test (3) with the first $I = 6$ curves. The bootstrap test statistic is $T = 0.16$, less than the critical value $\hat{T}_{0;0.01} = 0.24$. The $F$ test statistic is $F = 1.5$, also less than the critical value $F_{10;87;0.01} = 2.5$. The AIC difference is $\Delta = -3.84$. Consequently, the three procedures agree that the six curves are realizations from the same Michaelis-Menten model.

## Conclusions

FRET and/or BRET are techniques widely used to study protein-protein interactions in living cells. To evaluate these interactions, we used sample information to estimate the Michaelis-Menten parameters, $B_{max}$ and $K_d$. Here we considered and compared three ways of contrasting the homogeneity of several statistical samples obtained from FRET efficiency curves. Our aim was to determine whether $I > 1$ realizations of FRET experiments are homogeneous, in the sense that they are samples from a common underlying regression model. We focused on the Michaelis-Menten nonlinear regression model, since it is the most commonly used to fit this type of data, but the ideas can be extended to any other regression model. From the hypothesis testing point of view, we considered two test procedures for the null hypothesis of homogeneity, one based on bootstrap resampling and the other, a version of the classical $F$ test. We also used the AIC to decide which of these models (under the null or the alternative hypothesis) best fitted the data.

Observations from homogenous statistical samples obtained in the same experimental conditions can be pooled into a single sample. Once there is just one sample from each of two experiments performed in different experimental conditions, we might also use any of the statistical procedures considered here to determine whether there are differences due to the change in conditions.

A simulation study and analysis of real FRET data showed that the three methods used to study the homogeneity of FRET curves usually lead to the same conclusions. This, and the short time required to execute the program, suggests that for the analysis of real FRET saturation curves, it is feasible to use all three testing methods and verify that they lead to similar conclusions on sample homogeneity. It should nonetheless be taken into account that selection with AIC gave the best results for small sample sizes, whereas the F-test and bootstrap method should be selected for comparison of large samples. The Matlab code to implement the procedures described has been developed and tested by the authors.

## Authors' contributions

MM suggested the need for a homogeneity test. AB proposed the testing procedures, implemented them in Matlab and performed the computational experiments. LMM carried out the laboratory experiments and performed the homogeneity tests on the real data. MM,

LMM and AB prepared the manuscript. All authors analyzed the results, read and approved the final document.

## Acknowledgements

This work was partially supported by the Fundación Genoma España (MEICA), the European Union (FP7 Integrated Project Masterswitch no. 223404), the Spanish Ministry of Science and Innovation (SAF2008-02175 and MTM2010-17366), the DGUI de la Comunidad de Madrid/Universidad Autónoma de Madrid (CCG10-UAM/ESP-5494) and by the RETICS Program (RD08/0075 RIER; RD07/0020) of the Spanish Instituto de Salud Carlos III (ISCIII).

# Figures

**Figure 1. Samples of the saturation curve for CXCR4-CXCR4 dimer data**

Seven samples of the FRET saturation curve for CXCR4 dimer complexes in living cells. Each sample was obtained using HEK293T cells transiently cotransfected with a constant amount of CXCR4-CFP (2.0 µg, 4000 fluorescent unit [FU]) and increasing amounts of CXCR4-YFP (500-10,000 FU). Each sample corresponds to an individual experiment and is represented in a different color. The possible outliers have not yet been removed.

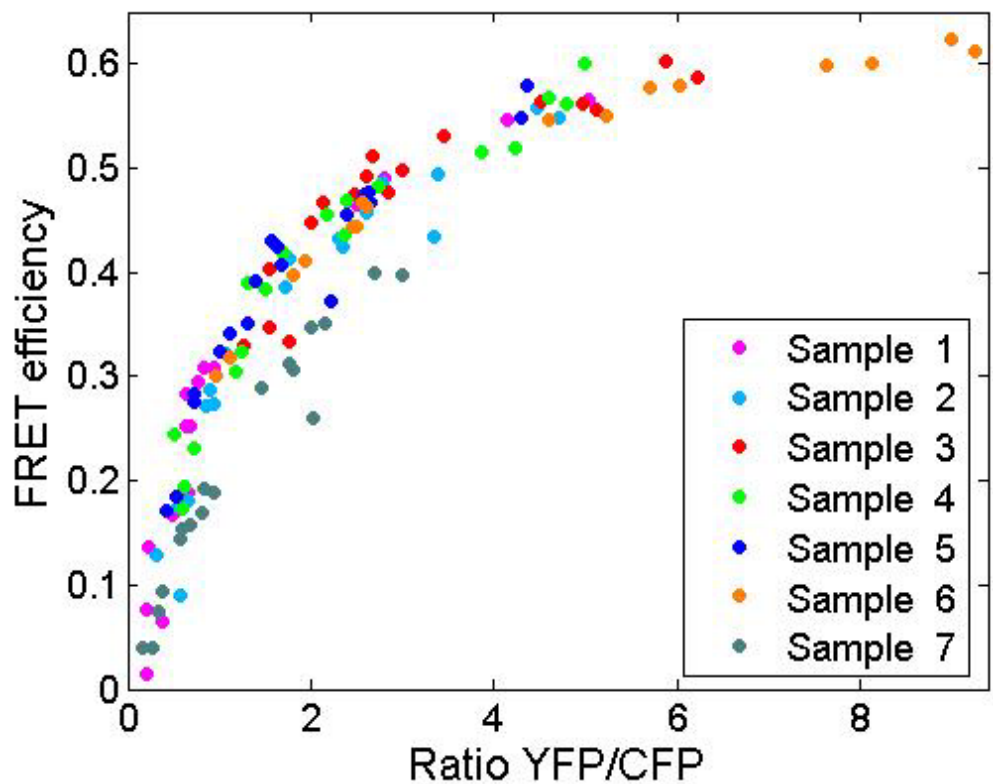

**Figure 2. Michaelis-Menten curves for CXCR4-CXCR4 dimer data**

The figure shows the estimated Michaelis-Menten curves for the data in Fig. 1. The parameters $B_{\max;i}$ and $K_{d;i}$ were estimated by maximum likelihood after removing the outliers. The estimates $\hat{B}_{\max;i}$ and $\hat{K}_{d;i}$ are marked on the vertical and horizontal axis, respectively.

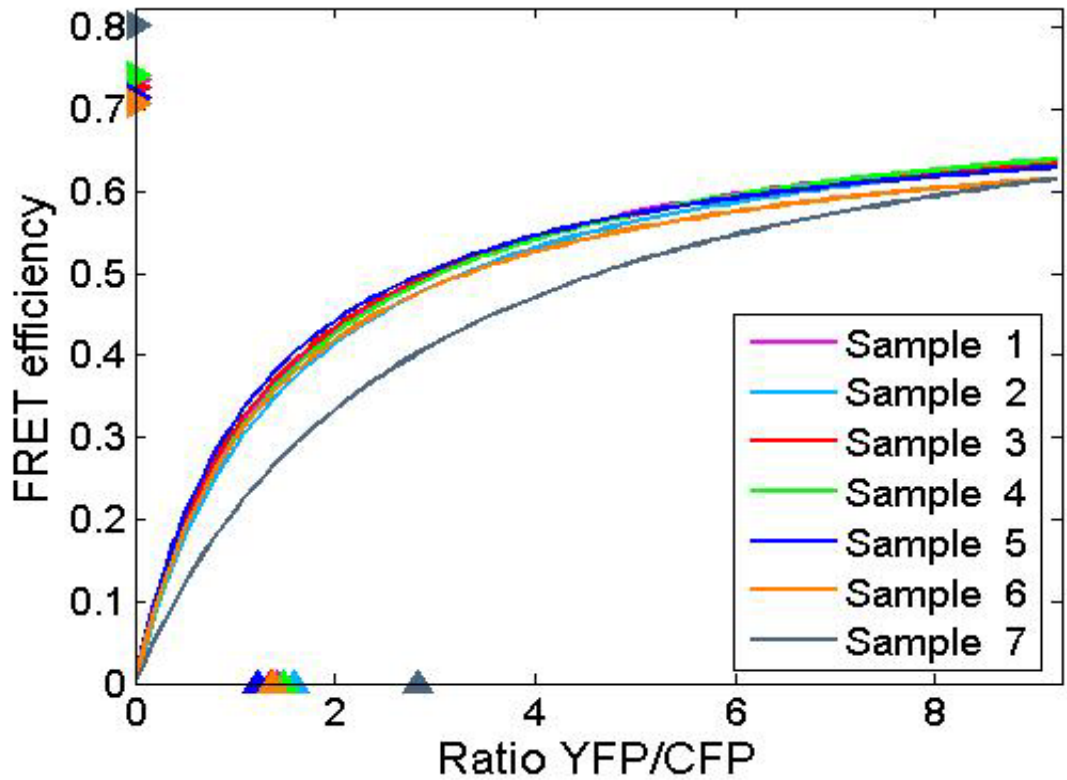

# Tables

## Table 1. Simulation results when $H_0$ is true

We display the proportion of $H_0$ rejections under the null hypothesis using the bootstrap procedure, the F test and the AIC (see Methods).

| $B_{max}$ | $K_d$ | $\sigma$ | $I$ | $n_i$ | Bootstrap | $F$ test | $\Delta \geq 2$ |
|---|---|---|---|---|---|---|---|
| 0.75 | 0.5 | 0.01 | 2 | 20 | 0.048 | 0.045 | 0.066 |
| | | | | 50 | 0.050 | 0.051 | 0.055 |
| | | | | 100 | 0.049 | 0.050 | 0.052 |
| | | | 5 | 20 | 0.051 | 0.052 | 0.033 |
| | | | | 50 | 0.049 | 0.050 | 0.024 |
| | | | | 100 | 0.054 | 0.047 | 0.027 |
| | | 0.001 | 2 | 20 | 0.040 | 0.039 | 0.063 |
| | | | | 50 | 0.048 | 0.047 | 0.058 |
| | | | | 100 | 0.058 | 0.057 | 0.060 |
| | | | 5 | 20 | 0.063 | 0.057 | 0.040 |
| | | | | 50 | 0.060 | 0.059 | 0.034 |
| | | | | 100 | 0.053 | 0.051 | 0.020 |
| 1 | 1.5 | 0.01 | 2 | 20 | 0.031 | 0.021 | 0.029 |
| | | | | 50 | 0.033 | 0.018 | 0.022 |
| | | | | 100 | 0.036 | 0.023 | 0.023 |
| | | | 5 | 20 | 0.043 | 0.015 | 0.010 |
| | | | | 50 | 0.051 | 0.023 | 0.010 |
| | | | | 100 | 0.038 | 0.008 | 0.003 |
| | | 0.001 | 2 | 20 | 0.023 | 0.012 | 0.015 |
| | | | | 50 | 0.035 | 0.023 | 0.026 |
| | | | | 100 | 0.027 | 0.016 | 0.016 |
| | | | 5 | 20 | 0.032 | 0.010 | 0.006 |
| | | | | 50 | 0.020 | 0.007 | 0.004 |
| | | | | 100 | 0.035 | 0.006 | 0.001 |

**Table 2. Simulation results when $H_0$ is false, for *I* = 2 curves**

We display the proportion of $H_0$ rejections under the alternative hypothesis using the bootstrap procedure, the F test and the AIC.

| $B_{max;1}$ | $B_{max;2}$ | $K_{d;1}$ | $K_{d;2}$ | $\sigma$ | $n_i$ | Bootstrap | $F$ test | $\Delta \geq 2$ |
|---|---|---|---|---|---|---|---|---|
| 0.75 | 0.76 | 0.50 | 0.50 | 0.01 | 20 | 0.236 | 0.235 | 0.285 |
|  |  |  |  |  | 50 | 0.560 | 0.553 | 0.571 |
|  |  |  |  |  | 100 | 0.990 | 0.990 | 0.992 |
| 0.75 | 0.75 | 0.50 | 0.51 | 0.01 | 20 | 0.109 | 0.104 | 0.130 |
|  |  |  |  |  | 50 | 0.247 | 0.244 | 0.253 |
|  |  |  |  |  | 100 | 0.496 | 0.489 | 0.498 |
| 1 | 1 | 1.50 | 1.51 | 0.01 | 20 | 0.055 | 0.033 | 0.048 |
|  |  |  |  |  | 50 | 0.090 | 0.051 | 0.059 |
|  |  |  |  |  | 100 | 0.121 | 0.084 | 0.097 |
| 0.99 | 1 | 1.50 | 1.50 | 0.01 | 20 | 0.270 | 0.203 | 0.242 |
|  |  |  |  |  | 50 | 0.596 | 0.531 | 0.550 |
|  |  |  |  |  | 100 | 0.977 | 0.974 | 0.974 |

### Table 3. Simulation results when $H_0$ is false, for $I = 3$ curves

We display the proportion of $H_0$ rejections under the alternative hypothesis using the bootstrap procedure, the F test and the AIC.

| $B_{max;1}$ | $B_{max;2}$ | $B_{max;3}$ | $K_{d;1}$ | $K_{d;2}$ | $K_{d;3}$ | $\sigma$ | $n_i$ | Bootstrap | F test | $\Delta \geq 2$ |
|---|---|---|---|---|---|---|---|---|---|---|
| 0.75 | 0.75 | 0.75 | 0.50 | 0.50 | 0.51 | 0.01 | 20 | 0.134 | 0.136 | 0.148 |
|  |  |  |  |  |  |  | 50 | 0.243 | 0.242 | 0.234 |
|  |  |  |  |  |  |  | 100 | 0.473 | 0.475 | 0.450 |
| 0.75 | 0.75 | 0.75 | 0.50 | 0.50 | 0.52 | 0.01 | 20 | 0.391 | 0.391 | 0.416 |
|  |  |  |  |  |  |  | 50 | 0.794 | 0.792 | 0.784 |
|  |  |  |  |  |  |  | 100 | 0.983 | 0.985 | 0.982 |
| 0.75 | 0.75 | 0.75 | 0.50 | 0.51 | 0.52 | 0.01 | 20 | 0.262 | 0.261 | 0.276 |
|  |  |  |  |  |  |  | 50 | 0.645 | 0.650 | 0.636 |
|  |  |  |  |  |  |  | 100 | 0.945 | 0.945 | 0.937 |
| 0.75 | 0.75 | 0.76 | 0.50 | 0.50 | 0.50 | 0.01 | 20 | 0.359 | 0.356 | 0.387 |
|  |  |  |  |  |  |  | 50 | 0.839 | 0.832 | 0.825 |
|  |  |  |  |  |  |  | 100 | 0.985 | 0.985 | 0.980 |
| 0.74 | 0.75 | 0.76 | 0.50 | 0.50 | 0.50 | 0.01 | 20 | 0.869 | 0.870 | 0.879 |
|  |  |  |  |  |  |  | 50 | 0.999 | 0.999 | 0.999 |
|  |  |  |  |  |  |  | 100 | 1.000 | 1.000 | 1.000 |
| 1 | 1 | 1 | 1.50 | 1.50 | 1.52 | 0.01 | 20 | 0.094 | 0.048 | 0.046 |
|  |  |  |  |  |  |  | 50 | 0.195 | 0.126 | 0.121 |
|  |  |  |  |  |  |  | 100 | 0.445 | 0.324 | 0.312 |
| 0.99 | 1 | 1 | 1.50 | 1.50 | 1.50 | 0.01 | 20 | 0.403 | 0.293 | 0.319 |
|  |  |  |  |  |  |  | 50 | 0.841 | 0.770 | 0.761 |
|  |  |  |  |  |  |  | 100 | 0.903 | 0.868 | 0.862 |

**Table 4. Summary of the FRET data**

We show sample sizes and estimated parameters in the Michaelis-Menten model for the FRET data in Fig. 1.

|  | Curves | | | | | | |
|---|---|---|---|---|---|---|---|
|  | 1 | 2 | 3 | 4 | 5 | 6 | 7 |
| $n_i$ | 16 | 18 | 17 | 18 | 18 | 18 | 18 |
| $\hat{B}_{max;i}$ | 0.7373 | 0.7417 | 0.7272 | 0.7424 | 0.7140 | 0.7079 | 0.8040 |
| $\hat{K}_{d;i}$ | 1.4203 | 1.5819 | 1.3512 | 1.4821 | 1.2329 | 1.3821 | 2.8197 |
| $\hat{\sigma}_i$ | 0.0389 | 0.0159 | 0.0181 | 0.0250 | 0.0168 | 0.0068 | 0.0124 |